\documentclass[12pt]{article}
\usepackage{amsfonts,amsmath}
\usepackage{amssymb}
\usepackage{ytableau}
\usepackage{dsfont}
\usepackage{hyperref}

\makeatletter \@addtoreset{equation}{section} \makeatother

\begin{document}
\begin{flushright}
%\texttt{Started on May 14}
%\\
{\small FIAN/TD/1-24}
\end{flushright}
\vspace{1.7 cm}
	
	\begin{center}
		{\large\bf Off-Shell Fields and Conserved Currents}
		
		\vspace{1 cm}
		
		{\bf E.O.~Spirin$^{1,2}$ and M.A.~Vasiliev$^{1,2}$}\\
		\vspace{0.5 cm}
		{\it
			$^1$I.E. Tamm Department of Theoretical Physics, Lebedev Physical Institute,\\
			Leninsky prospect 53, 119991, Moscow, Russia}\\
		\vspace{0.5 cm}
		{\it
			$^2$Moscow Institute of Physics and Technology,\\
			Institutsky lane 9, 141700, Dolgoprudny, Moscow region, Russia}
	\end{center}

	\begin{abstract}
		We study interactions of higher-spin massless fields $\varphi$ with conserved currents multilinear in the off-shell matter fields $\phi$. Specifically, we focus on the 3d case where a slight modification of
the  $\sigma_-$-cohomology technique developed earlier is directly applicable to control nontriviality of the interaction vertices at the convention that the vertices removable by a local field redefinition of a higher-spin field or having  schematic form $F(\varphi) G(\phi)$, where $F(\varphi)$ is a gauge invariant field strength of a free higher-spin field, are called deformationally trivial. It is demonstrated how the $\sigma_-$-cohomology approach can be applied to the analysis of nonlinear vertices.
Generally, deformationally trivial vertices are not $\sigma_-$-closed while the deformationally non-trivial ones must be in $H(\sigma_-)$. It is shown that, at least in the $3d$
case, the relevant cohomology group $H^1(\sigma_-)=0$ and, hence, no deformationally non-trivial
off-shell vertices exist. On the other hand, there exists an infinite class of deformationally trivial vertices, that includes the vertex recently proposed for spin three. Our analysis goes beyond the higher-spin vertices
allowing to show that, at least in three dimensions, nonlinear combinations of the
off-shell scalar fields and their derivatives cannot obey  non-trivial equations.
			\end{abstract}
\newpage
%\textheight 22.6 true cm
\tableofcontents
\newpage	

	\section{Introduction} \label{sec1}

In the Fronsdal theory of massless fields \cite{Fronsdal},  a
spin--$s$ field $\varphi_{\mu_1 \dots \mu_s}(x)$ obeys the double-tracelessness conditions
$\eta^{\alpha \beta}\eta^{\gamma \delta}\varphi_{\alpha \beta \gamma \delta \mu_5\ldots \mu_s}=0$.
Spin--$s$ massless field equations
\begin{equation}
{\cal R}_{\mu_1 \dots \mu_s}(\varphi):=	\Box \varphi_{\mu_1 \dots \mu_s} - s \partial_{(\mu_1}\partial^\alpha \varphi_{\mu_2 \dots \mu_s)\alpha} + \frac{s(s-1)}{2} \partial_{(\mu_1} \partial_{\mu_2} \varphi_{\mu_3\dots \mu_s) \alpha}^{\;\;\;\;\;\;\;\;\;\;\;\;\;\;\alpha} = 0\,,
\end{equation}
where $\partial_\mu:=\frac{\partial}{\partial x^\mu}$,
are invariant under the  gauge transformations
\begin{equation}\label{gauge}
	\delta \varphi_{\mu_1 \dots \mu_s}(x) = \partial_{(\mu_1} \xi_{\mu_2 \ldots \mu_s)}(x)
\end{equation}
with traceless gauge parameters $\eta^{\nu\rho}\xi_{\nu\rho \mu_4\ldots \mu_s}(x)=0$.

It is well known how to construct interactions of higher-spin (HS)
massless fields with conserved currents $J(\phi)$ bilinear in the fluctuating fields
$\phi$ of various spins (see e.g. \cite{Berends:1985xx,Metsaev:2007rn, Gelfond:2013lba}).
In that case, the HS currents
$J(\phi)$ are conserved on shell, {\it i.e.} at the conditions that the fields $\phi$ obey
their free field equations.

In the paper \cite{Lavrov}  a conserved spin-three current
 $ T^{\mu \nu \lambda}[\phi]$
 trilinear in the  off-shell scalar
 field $\phi$ was found within the
BV-formalism.
An off-shell field is a field that obeys no equations of motion. Nevertheless, multilinear combinations of derivatives of such fields, which we call composite  fields,
still can satisfy some partial
differential equations. For instance, such equations may be associated
with  the gauge invariance. In particular,  equations for
the spin-three current $T^{\mu\nu\lambda}$, built from an off-shell  scalar  $\phi$,
\begin{equation*}
	T^{\mu \nu \lambda} =\big[ \partial^\mu \partial^\nu \partial^\lambda \phi \; \phi^2 + 2 \partial^{(\mu} \partial^\nu \phi \; \partial^{\lambda)} \phi \; \phi + 2 \partial^\mu \phi \; \partial^\nu \phi \; \partial^\lambda \phi - \frac{1}{2} \eta ^{(\mu \nu} \partial^{\lambda)} \Box \phi \; \phi^2 -
\end{equation*}
\begin{equation} \label{current}
	-2 \eta^{(\mu \nu} \partial^{\lambda)} \partial_\sigma \phi \; \partial^\sigma \phi \; \phi - \eta^{(\mu \nu} \partial^{\lambda)} \phi \; \partial_\sigma \phi \; \partial^\sigma \phi - \Box \phi \; \eta^{(\mu \nu} \partial^{\lambda)} \phi \; \phi \big]\,,
\end{equation}
found in \cite{Lavrov}\footnote{Here (...) denotes the total symmetrization:$A_{(\mu_1 \dots \mu_k)} = \frac{1}{k!} \sum_{\sigma} A_{\sigma(\mu_1)\dots \sigma(\mu_k)}$.}, imply the gauge invariance of the quartic spin-three spin-zero vertex
 \begin{equation}
\label{vert}
S=\int d^dx \; \varphi_{\mu\nu\lambda} T^{\mu \nu \lambda}[\phi]
\end{equation}
under the gauge transformations
\begin{equation}
	\delta \varphi^{\mu \nu \lambda} = \partial^{(\mu} \xi^{\nu \lambda)}, \;\;\;\;\; \delta \phi = 0, \;\;\;\;\; \eta_{\mu \nu} \xi^{\mu \nu} = 0,
\end{equation}
where $\xi^{\mu \nu}$ is a traceless gauge parameter, $\Box = \partial_\mu \partial^\mu$, and $\eta_{\mu \nu}$ is the Minkowski metric in $d$ dimensions.
This implies the current conservation equation
\begin{equation}
\label{ccon}
	\xi_{\nu \lambda} \partial_\mu T^{\mu \nu \lambda} = 0 \quad \Rightarrow
\quad \partial_\mu T^{\mu \nu \lambda} = \eta^{\nu \lambda} K[\phi]
\end{equation}
with some $K[\phi]$.

Note that, though in our case the field $\phi$ is off-shell, the same gauge-invariant vertex exists for an on-shell scalar field with the standard kinetic term in the Lagrangian.
Analogous interaction vertices both in Minkowski  and AdS space were studied more generally in
\cite{Fredenhagen:2019hvb, Joung:2019wbl, Fredenhagen:2019lsz}.
According to  these works such  vertices must be either  on-shell
trivial ({\it i.e.}, removable by a local field redefinition), or composed from the
gauge invariant field strengths in which case the gauge transformations remain
undeformed (Abelian). Both types of such vertices will be called deformationally
trivial to stress that they do not induce a deformation of the HS gauge symmetry. Note that
cubic vertices for arbitrary spins in $3d$ Minkowski space were classified in \cite{Metsaev:2020gmb}.

In this paper we develop a technics allowing to check whether there exist
composite fields built from the off-shell fields, that obey
some non-trivial partial differential equations including the conservation conditions
like (\ref{ccon}).
This problem will be analyzed in terms of the
$\sigma_-$-cohomology technique \cite{Shaynkman:2000ts} allowing
  to control the dynamical content of the theory in question. Note that this approach
  is free from the additional conjectures on the possibility to
   use non-local inverse operators as in
  \cite{Fredenhagen:2019hvb, Joung:2019wbl, Fredenhagen:2019lsz}.

 Those components $T^{pr}(x)$  of the (composite) fields,
 that cannot be expressed  via derivatives of the other components, are called dynamical (primary). Primary components
 of the fields described by the $p$-forms are
 represented by the $\sigma_-$-cohomology $H^p(\sigma_-)$ \cite{Shaynkman:2000ts}.
 The primary components may or may not obey some differential equations represented
 by $H^{p+1}(\sigma_-)$. The descendants $T^{dec}=L(\partial) T^{pr}$ result from the action of various differential operators on the primary components. There are
 always some descendant fields that obey certain differential equations (e.g.,
  Bianchi identities) expressing the properties of the operators $L(\partial)$.
  Such equations are in a certain sense trivial expressing the properties of
  $L(\partial)$ rather than  $T^{pr}(x)$. These are not represented by any $\sigma_-$-cohomology. Specifically, in the HS theory, there are two types of the operators $L(\partial)$ leading to the
  gauge invariant current interactions. These are  $L_F(\partial)$ associated with
  the l.h.s. of the Fronsdal equations and $L_W(\partial)$ associated with the
  gauge invariant combinations of the Fronsdal fields like Faradey tensor in the spin-one
  case or Weyl tensor in the spin-two case. The terms with $L_F(\partial)$ are removable
  by appropriate field redefinitions in the free action. Those with $L_W(\partial)$
 describe interactions with the gauge invariant field strengths. These
  are not removable by a local field redefinition but do not affect the gauge
  transformation law, hence called deformationally trivial.

Specifically, the $\sigma_-$-cohomology analysis of the so-called higher-rank fields
to be used in this paper, has been elaborated in  \cite{Gelfond:2013lba} for the $3d$ case.
In these terms, a $3d$ off-shell scalar field can be realized as a rank-two field while its
cubic combination is a rank-six field. However, the approach developed in
\cite{Gelfond:2013lba} for fields of any rank operates  with the lowest weight vectors
of the Howe-dual algebra (for more detail see Section \ref{sec5}) while, as discussed in
Section \ref{sec4}, the off-shell scalar fields carry zero weights, that requires some modification
of the scheme. Our final result shows that the $\sigma_-$-cohomology representing
possible non-trivial partial differential equations (in particular, conservation conditions)  obeyed by cubic combinations
$ T^{\mu \nu \lambda}[\phi]$ of the $3d$ off-shell scalar fields $\phi$ is empty.
In turn, in agreement with \cite{Fredenhagen:2019hvb, Joung:2019wbl, Fredenhagen:2019lsz},
this implies that $ T^{\mu \nu \lambda}[\phi]$ cannot be a primary conserved
current. Hence, it has to be an improvement while the vertex  (\ref{vert}) must be
removable by a local field redefinition. The explicit form of the latter is found
in Section \ref{sec6}.

The aim of this paper is to illustrate how the $\sigma_-$-cohomology technique
can be used for the analysis of (non-)triviality of vertices. As a  consequence
it will be  shown that, at least in three space-time
dimensions, there are no primary composite fields, that obey non-trivial equations
(including, the conservation conditions) built from
any number of the off-shell scalar fields and their derivatives as well as from HS conserved currents.

The rest of the paper is organized as follows. The $\sigma_-$-cohomology technique is recalled in Section \ref{sec3}. In Sections \ref{sec4} and \ref{sec5}, the analysis of the $\sigma_-$-cohomology
groups $H^0(\sigma_-)$ and $H^1(\sigma_-)$ is presented for the composite
fields built from the $3d$ off-shell scalar fields and HS conserved currents.
In Section \ref{sec6}, our results are compared against those of \cite{Lavrov}. Section \ref{sec7}
contains conclusions.  Some technical details of the
$\sigma_-$-cohomology analysis are in Appendix.

\section{$\sigma_-$-cohomology} \label{sec3}

In this section, we  briefly sketch  relevant aspects of
the HS theory and $\sigma_-$-cohomology approach
referring for more detail
to \cite{Bekaert:2004qos}.

The $Sp(2M)$-invariant equations, that describe  $3d$ and $4d$ massless fields at $M=2$ and $M=4$, respectively, are  \cite{Vasiliev:2001zy}
\begin{equation} \label{3.1}
	\left(dX^{\alpha \beta} \frac{\partial}{\partial X^{\alpha \beta}} + \sigma_- \right)C(y|X) = 0, \;\;\;\;\; \sigma_- = dX^{\alpha \beta} \frac{\partial^2}{\partial y^\alpha \partial y^\beta}, \;\;\;\;\; (\sigma_-)^2=0\,,
\end{equation}
where $y^\alpha$ are auxiliary commuting variables,
\begin{equation}
\label{rank1}
	C(y|X)= \sum_{n=0}^{\infty} C_{\alpha_1 \dots \alpha_n}(X) y^{\alpha_1} \dots y^{\alpha_n}\,,
\end{equation}
 $dX^{\alpha \beta}$ are anticommuting differentials for symmetric matrix space-time coordinates $X^{\alpha \beta} = X^{\beta \alpha}$ and $\alpha, \beta = 1,\dots,M$. In this paper we confine ourselves to the case of
 $M=2$ associated with a $3d$ massless theory.

Let us recall some  terminology. A field is called auxiliary if it is expressed in terms of other fields and their derivatives by virtue of equations of motion or can be gauge fixed to
zero by some shift (Stueckelberg) gauge symmetry. The dynamical sector consists  of the fields that are neither expressed in terms of the other fields nor are pure gauge. Thus, it makes sense to focus on the dynamical fields and their equations of motion. For instance, Fronsdal fields are of that type.

It can be shown \cite{Shaynkman:2000ts} that dynamical fields and their field equations associated with the equations \eqref{3.1} are contained in zeroth cohomology $H^0(\sigma_-)$ and first cohomology $H^1(\sigma_-)$, respectively. Recall that
\begin{equation}
	H^p(\sigma_-) := \frac{\ker (\sigma_-)}{\text{Im} (\sigma_-)}\bigg|_p,
\end{equation}
where $|_p$ implies restriction to the space of $p-$forms. That dynamical fields are in zeroth cohomology is fairly simple. Indeed, it is  obvious from (\ref{3.1}) that the fields, that are not in $\ker \sigma_-$, are expressed  in terms of space-time derivatives of the fields from  $\ker \sigma_-$.

The fields (\ref{rank1}) describe a so-called rank-one system.
The rank-$r$ fields are
\begin{equation}
	C(y|X) = \sum_{n} f^{i_1 \dots i_n}_{\alpha_1 \dots \alpha_n}(X) y_{i_1}^{\alpha_1} \dots y_{i_n}^{\alpha_n},
\end{equation}
where $i_k = 1 \dots, r$. The rank-two system  with
\begin{equation} \label{r2}
(\sigma_-)_{r=2} =  \frac{\partial}{\partial y_{-}^\alpha} \frac{\partial}{\partial y_{+}^\beta} dX^{\alpha \beta}
\end{equation}
was studied in \cite{Gelfond:2003vh} where  it was shown, in particular, that if the operator
\begin{equation}
	  h = y_-^\alpha \frac{\partial}{\partial y_-^\alpha} - y_+^\alpha \frac{\partial}{\partial y_+^\alpha}, \quad [h,(\sigma_-)_{r=2}] = 0
\end{equation}
has  zero eigenvalue then the corresponding fields are off-shell scalars. We will consider them as off-shell fields from which the current (\ref{current}) is built.

In our case the relevant operator $\sigma_-$ is
\begin{equation}
\label{gs-}
	\sigma_- = \sum_{i} \frac{\partial}{\partial y_{-,i}^\alpha} \frac{\partial}{\partial y_{+,i}^\beta} dX^{\alpha \beta}.
\end{equation}

We are interested in the trilinear combinations of scalar off-shell fields because the  current (\ref{current}) is cubic in them. Each of the three pairs of variables $y_{\pm,i}$ is associated with its own scalar field, that in addition obeys the
restriction
\begin{equation}\label{H}
h_j C(y_j|X) =  0\,,\qquad h_j:= \left(y_{-,j}^\alpha \frac{\partial}{\partial y_{-,j}^\alpha}-y_{+,j}^\alpha\frac{\partial}{\partial y_{+,j}^\alpha}\right)\,,\qquad \forall j\,.
\end{equation}
This is the same restriction as (\ref{r2}) in the rank-two case.

Since the operators $h_j$ commute with $\sigma_-$,
\begin{equation}
	[h_j,\sigma_-] = 0\,,
\end{equation}
the fields corresponding to the different eigenvalues of $h_j$ form different representations. Hence, each eigenvalue corresponds to its own subsystem. In  \cite{Gelfond:2003vh} it was shown that, generally, such systems describe conserved currents of various spins, that include currents of spin zero satisfying no conservation conditions
thus being off-shell. So, we analyze a three-linear current built from the spin zero fields, that obey \eqref{H}.

In the end of this section let us recall
 the Poincar\'e homotopy trick conventionally used to calculate the cohomology group. Let $D$ be a linear operator acting in a Hilbert space $V$, such that $D^2=0$. Let there also be a nilpotent operator $D^*$, $(D^*)^2=0$. The homotopy operator is
\begin{equation}
	\Delta := \left\{ D,D^* \right\}\,.
\end{equation}
By its definition it obeys
\begin{equation}
	[D,\Delta] = [D^*, \Delta] = 0\,
\end{equation}
   and, hence,
   $$\Delta \ker D \subset \text{Im} D\,.$$
   Then it follows that $H(D) := \ker D / \text{Im} D \subset \ker D / \Delta (\ker D)$. Let the operators $D$ and $D^*$ be conjugated in the Hilbert space $V$ and $V=\sum \oplus V_A$ for a set of finite-dimensional $V_A$ such
   that $\Delta(V_A) \subset V_A$ and $V_A$ be orthogonal to $V_B$ at $A \neq B$. Then $\Delta$ can be diagonalized after which it is easily observed that $\ker D /\Delta(\ker D) = \ker \Delta \cap \ker D$. As a result,
\begin{equation} \label{cohD}
	H(D) \subset \ker \Delta \cap \ker D\,
\end{equation}
and the problem amounts to   solution of the homotopy equation
\begin{equation} \label{homotopy}
	\Delta f = 0\,.
\end{equation}
As shown in \cite{Gelfond:2013lba} for $H^p(\sigma_-)$ with any $p$ formula (\ref{cohD}) is an equality, $H(D) = \ker \Delta \cap \ker D$.

\section{$H^0(\sigma_-$)} \label{sec4}
Here we analyze $H^0(\sigma_-)$ with $\sigma_-$ (\ref{gs-}), where the
index  $i$ ranges from 1 to 3 because we are interested in the cubic combinations of the
off-shell scalar fields. The variables $y_{\pm,i}$ with different $i$ are
associated with different off-shell scalar fields.
Let us  rewrite $\sigma_-$ in the form
\begin{equation}\label{color}
	\sigma_- =  \frac{\partial}{\partial y_a^\alpha} \frac{\partial}{\partial y_b^\beta} \eta_{a b} dX^{\alpha \beta}, \;\;\;\;\; \eta_{ab} = \left(
	\begin{array}{cc}
		0 & \mathds{1}_{3 \times 3} \\
		\mathds{1}_{3 \times 3} & 0
	\end{array}
	\right)\,,\qquad a,b =1\ldots 6\,,
\end{equation}
where indices are relabeled as follows:
$[(-,1);(-,2);(-,3);(+,1);(+,2);(+,3)] \rightarrow [1,2,3,4,5,6]$.

Recall that the scalar fields are those with $\# y_+=\# y_-$ which condition
is expressed by  (\ref{H}).
The operators $h_j$ (\ref{H}) belong to $o(3,3)$ with the generators
$\tau_{ij}$,
\begin{equation}
	\tau_{mn} = y_{m}^{\alpha} \frac{\partial}{\partial y^{\alpha,n}} - y_n^{\alpha} \frac{\partial}{\partial y^{\alpha,m}}, \;\;\;\;\; [\tau_{mn},\tau_{kl}]= \eta_{nk}\tau_{ml}+\eta_{ml}\tau_{nk}-\eta_{mk}\tau_{nl}-\eta_{nl}\tau_{mk} \,,
\end{equation}
that commute with $\sigma_-$,
\begin{equation}
[\sigma_-,\tau_{mn}]=0\,.
\end{equation}
In the original index notations $\tau_{mn}$ read as
\begin{equation}
	\tau_{12}=y_{-,1}^\alpha \frac{\partial}{\partial y_{+,2}^\alpha}-y_{-,2}^\alpha\frac{\partial}{\partial y_{+,1}^\alpha}, \;\;\;\;\; \tau_{13}=y_{-,1}^\alpha \frac{\partial}{\partial y_{+,3}^\alpha}-y_{-,3}^\alpha \frac{\partial}{\partial y_{+,1}^\alpha},
\end{equation}
\begin{equation}
	\tau_{14}=y_{-,1}^\alpha \frac{\partial}{\partial y_{-,1}^\alpha}-y_{+,1}^\alpha\frac{\partial}{\partial y_{+,1}^\alpha}, \;\;\;\;\; \tau_{15}=y_{-,1}^\alpha \frac{\partial}{\partial y_{-,2}^\alpha}-y_{+,2}^\alpha \frac{\partial}{\partial y_{+,1}^\alpha},
\end{equation}
\begin{equation}
	\tau_{16}=y_{-,1}^\alpha \frac{\partial}{\partial y_{-,3}^\alpha}-y_{+,3}^\alpha\frac{\partial}{\partial y_{+,1}^\alpha}, \;\;\;\;\; \tau_{23}=y_{-,2}^\alpha \frac{\partial}{\partial y_{+,3}^\alpha}-y_{-,3}^\alpha \frac{\partial}{\partial y_{+,2}^\alpha},
\end{equation}
\begin{equation}
	\tau_{24}=y_{-,2}^\alpha \frac{\partial}{\partial y_{-,1}^\alpha}-y_{+,1}^\alpha\frac{\partial}{\partial y_{+,2}^\alpha}, \;\;\;\;\; \tau_{25}=y_{-,2}^\alpha \frac{\partial}{\partial y_{-,2}^\alpha}-y_{+,2}^\alpha \frac{\partial}{\partial y_{+,2}^\alpha},
\end{equation}
\begin{equation}
	\tau_{26}=y_{-,2}^\alpha \frac{\partial}{\partial y_{-,3}^\alpha}-y_{+,3}^\alpha\frac{\partial}{\partial y_{+,2}^\alpha}, \;\;\;\;\; \tau_{34}=y_{-,3}^\alpha \frac{\partial}{\partial y_{-,1}^\alpha}-y_{+,1}^\alpha \frac{\partial}{\partial y_{+,3}^\alpha},
\end{equation}
\begin{equation}
	\tau_{35}=y_{-,3}^\alpha \frac{\partial}{\partial y_{-,2}^\alpha}-y_{+,2}^\alpha\frac{\partial}{\partial y_{+,3}^\alpha}, \;\;\;\;\; \tau_{36}=y_{-,3}^\alpha \frac{\partial}{\partial y_{-,3}^\alpha}-y_{+,3}^\alpha \frac{\partial}{\partial y_{+,3}^\alpha},
\end{equation}
\begin{equation}
	\tau_{45}=y_{+,1}^\alpha \frac{\partial}{\partial y_{-,2}^\alpha}-y_{+,2}^\alpha\frac{\partial}{\partial y_{-,1}^\alpha}, \;\;\;\;\; \tau_{46}=y_{+,1}^\alpha \frac{\partial}{\partial y_{-,3}^\alpha}-y_{+,3}^\alpha \frac{\partial}{\partial y_{-,1}^\alpha},
\end{equation}
\begin{equation}
	\tau_{56}=y_{+,2}^\alpha \frac{\partial}{\partial y_{-,3}^\alpha}-y_{+,3}^\alpha\frac{\partial}{\partial y_{-,2}^\alpha}
\end{equation}
with mutually commuting Cartan elements
\begin{equation}
	h_1=\tau_{14}\,,\qquad h_2 = \tau_{25}\,,\qquad h_3 = \tau_{36}\,.
\end{equation}
Introducing operators
$y_i^\alpha$ and $W_\alpha^i = \frac{\partial}{\partial y_i^\alpha}$, that obey
\begin{equation}
	[ W_\alpha^i, y_j^\beta ] = \delta_j^i \delta_\alpha^\beta, \;\;\;\;\; [ y_i^\alpha, y_j^\beta ] = 0, \;\;\;\;\; [ W_\alpha^i, W_\beta^i ] = 0\,,
\end{equation}
 $sp(2M)$ can be represented in the form
\begin{equation}
	T^{\alpha \beta} = y_i^\alpha y_j^\beta \delta^{ij}, \;\;\;\;\; T_{\alpha \beta} = W_\alpha^i W_\beta^j \delta_{ij}, \;\;\;\;\; T_\beta^\alpha = \frac{1}{2} \left\{ y_j^\alpha, W_\beta^j \right\},
\end{equation}
\begin{equation}
	[T_{\alpha \beta}, T^{\gamma \delta}] = \delta_\alpha^\gamma T_\beta^\delta + \delta_\alpha^\delta T_\beta^\gamma + \delta_\beta^\gamma T_\alpha^\delta + \delta_\beta^\delta T_\alpha^\gamma,
\end{equation}
\begin{equation}
	[T_\alpha^\beta, T^{\gamma \delta}] = \delta_\alpha^\gamma T^{\beta \delta} + \delta_\alpha^\delta T^{\beta \gamma}, \;\;\;\;\; [T_\alpha^\beta, T_{\gamma \delta}] = -\delta_\gamma^\beta T_{\alpha \delta} - \delta_\delta^\beta T_{\alpha \gamma}.
\end{equation}
Being mutually commuting, algebras $o(3,3)$ and $sp(2M)$ form a Howe dual pair \cite{Howe}.

The operator $\sigma_-^*$ conjugated to $\sigma_-$ is defined analogously to \cite{Gelfond:2013lba}:
\begin{equation}
		\sigma_- = T_{\alpha \beta} dX^{\alpha \beta}, \quad \sigma_-^* = T^{\alpha \beta} \frac{\partial}{\partial (dX^{\alpha \beta})}, \;\;\;\;\; (\sigma_-^*)^2 = 0\,.
\end{equation}
An elementary  computation yields
\begin{equation}
\label{De}
\Delta := \left\{ \sigma_-, \sigma_-^* \right\}= \nu^\alpha_\beta \nu^\beta_\alpha + \frac{1}{2} \tau_{mn}\tau^{mn} - (M - 2)\nu_\alpha^\alpha,
\end{equation}
where $\frac{1}{2}\tau_{mn}\tau^{mn}$ is the quadratic Casimir  of $o(3,3)$,
\begin{equation}
	\nu_\beta^\alpha := \rho^\alpha_\beta + \Upsilon^\alpha_\beta, \;\;\;\;\; \rho^\alpha_\beta := 2 dX^{\alpha \gamma} \frac{\partial}{\partial (dX^{\beta \gamma})}, \;\;\;\;\; \Upsilon^\alpha_\beta := T^\alpha_\beta - \frac{3}{2} \delta^\alpha_\beta\,,
\end{equation}
\begin{equation}
	\frac{\partial}{\partial (dX)^{\alpha \beta}} dX^{\gamma \delta} = \frac{1}{2} \left( \delta_\alpha^\gamma \delta_\beta^\delta + \delta_\alpha^\delta \delta_\beta^\gamma \right) - dX^{\gamma \delta} \frac{\partial}{\partial (dX)^{\alpha \beta}}\,.
\end{equation}

To calculate  the part of $H^0(\sigma_-)$ associated with the off-shell scalar fields
  we (i) choose a Chevalley basis in $o(3,3)$, consisting of raising $e_i$, lowering $f_i$ and Cartan operators $h_i$, (ii) find the lowest vectors annihilated by the lowering operators, (iii) act by
the raising operators on the lowest vectors to obtain  the vectors annihilated by the Cartan operators \begin{equation}\label{vac}
	f_1 |vac\rangle = f_2 |vac \rangle = f_3 |vac \rangle =0\,,
\end{equation}
\begin{equation} \label{hvac}
	h_1 F(e_1,e_2,e_3)|vac \rangle = h_2 F(e_1,e_2,e_3)|vac \rangle = h_3 F(e_1,e_2,e_3)|vac \rangle =  0.
\end{equation}

Once $ |vac \rangle\in \ker \Delta$, from the commutativity of $o(3,3)$ and $sp(2M)$
it follows that
any vector $F(e_1,e_2,e_3)|vac \rangle$ also belongs to $\ker \Delta$. Hence all
solutions of (\ref{hvac}) also belong to $H^0(\sigma_-)$.

The Chevalley decomposition is ($\alpha_i$ are simple roots)
\begin{equation}
\label{ch}
	\tilde{h}_i = \frac{2}{\langle \alpha_i \alpha_i \rangle} h_{\alpha_i}, \;\;\;\;\; e_i = e_{\alpha_i}, \;\;\;\;\; f_i = \frac{2}{\langle  \alpha_i \alpha_i \rangle (e_{\alpha_i}, e_{-\alpha_i})} e_{-\alpha_i}, \;\;\;\;\; (e_i, f_i) = \frac{2}{\langle \alpha_i \alpha_i \rangle}.
\end{equation}
In this basis the commutation relations are
\begin{equation}
	[\tilde{h}_i, \tilde{h}_j] = 0, \;\;\;\;\; [\tilde{h_i}, e_i] = a_{ij} e_j, \;\;\;\;\; [\tilde{h}_i, f_j] = - a_{ij} f_j, \;\;\;\;\; [e_i,f_j] = \delta_{ij} \tilde{h}_i,
\end{equation}
where $a_{ij}$ is the Cartan matrix,
\begin{equation}
	a_{ij} = \frac{2 \langle \alpha_j \alpha_i \rangle}{\langle \alpha_i \alpha_i \rangle}\,.
\end{equation}
In the case of interest it reads as
\begin{equation}
	a_{ij} = \left(
	\begin{array}{ccc}
		2 & -1 & -1 \\
		-1 & 2 & 0 \\
		-1 & 0 & 2
	\end{array}
	\right)
\end{equation}
and  the
Chevalley basis is
\begin{equation}\label{th}
	\tilde{h}_1 = h_1+h_2, \;\;\;\;\; \tilde{h}_2 = -h_1 + h_3, \;\;\;\;\; \tilde{h}_3 = -h_1-h_3,
\end{equation}
\begin{equation}
	e_1=\tau_{12}, \;\;\;\;\; e_2=\tau_{34}, \;\;\;\;\; e_3=\tau_{46}, \;\;\;\;\; f_1=-\tau_{45}, \;\;\;\;\; f_2=\tau_{16}, \;\;\;\;\; f_3=-\tau_{13}.
\end{equation}
It is worth mentioning that the zero-weight constraints (\ref{hvac}) are equivalent for $\tilde{h}_i$ and $h_i$.

Let us start with the lowest weight vacuum conditions (\ref{vac})
\begin{equation}\label{soe}
	\begin{cases}
		\left( y_{-,1}^\alpha \frac{\partial}{\partial y_{-,3}^\alpha}-y_{+,3}^\alpha\frac{\partial}{\partial y_{+,1}^\alpha} \right) C = 0, \\
		\left( y_{-,1}^\alpha \frac{\partial}{\partial y_{+,3}^\alpha}-y_{-,3}^\alpha \frac{\partial}{\partial y_{+,1}^\alpha} \right) C = 0,\\
		\left( y_{+,1}^\alpha \frac{\partial}{\partial y_{-,2}^\alpha}-y_{+,2}^\alpha\frac{\partial}{\partial y_{-,1}^\alpha} \right) C =0,\\
		\tilde{h_1} C = \mu_1 C, \;\; \tilde{h_2} C = \mu_2 C, \;\; \tilde{h_3} C = \mu_3 C\,,
	\end{cases}
\end{equation}
where $\mu_i$ are the lowest weights. The idea is to analyze the problem for general $\mu_i$, setting $\mu_i=0$ in the end.

According to the method  of characteristics, the general solution of a system of first-order equations is a function of elementary solutions. The elementary solutions
to the first three equations (\ref{soe}) are
\begin{equation}
	G_1^\alpha = y_{+,2}^\alpha, \quad G_2^{\alpha \alpha} = y_{+,1}^\alpha y_{-,1}^\alpha + y_{+,2}^\alpha y_{-,2}^\alpha + y_{-,3}^\alpha y_{+,3}^\alpha.
\end{equation}
The proof that these are indeed elementary solutions will be given in Appendix
\ref{Appendix}.

The solution $G_1$ is obvious because there are no $y_{+,2}^\alpha$ derivatives in
(\ref{soe}).
The solution $G_2$ is irrelevant because it does not belong to the kernel of $\sigma_-$ while $G_1$ does hence belonging  to $H^0(\sigma_-)$. Now we are in a position to analyze how the raising operators act on this solution. As  mentioned above, functions of elementary solutions also satisfy the set of equations. So we will take for example $\psi_1^{\alpha(2)} = y_{+,2}^{\alpha} y_{+,2}^{\alpha}$ and $\psi_2^{\alpha(4)} = y_{+,2}^{\alpha} y_{+,2}^{\alpha} y_{+,2}^{\alpha} y_{+,2}^{\alpha}$.
 Searching through all the possible combinations of raising operators, we find the solutions, annihilated by the Cartan operators:
\begin{equation}
	e_3 e_1 e_2 e_1 \psi_1^{\alpha(2)} = 2y_{+,1}^{\alpha}y_{-,1}^{\alpha}-2y_{-,3}^{\alpha}y_{+,3}^{\alpha}, \;\;\;\;\; e_1e_3e_2e_1\psi_1^{\alpha(2)} = 2y_{+,1}^{\alpha} y_{-,1}^{\alpha}-2y_{-,2}^{\alpha} y_{+,2}^{\alpha},
\end{equation}
\begin{equation}
	e_2e_1e_3e_1 \psi_1^{\alpha(2)} = 2y_{+,1}^{\alpha} y_{-,1}^{\alpha}-2y_{+,3}^{\alpha} y_{-,3}^{\alpha}.
\end{equation}
A linear combination of these solutions can be written as
\begin{equation}
	\Psi_1^{\alpha(2)} =Ay_{+,1}^{\alpha}y_{-,1}^{\alpha}+By_{-,2}^{\alpha} y_{+,2}^{\alpha}+Cy_{+,3}^{\alpha} y_{-,3}^{\alpha}, \;\;\;\;\; A+B+C=0.
\end{equation}
It is easy to check that $\Psi_1^{\alpha(2)}$ belongs to the kernel of operator $\sigma_-$. 
Now we  examine $\psi_2^{\alpha(4)}$:
\begin{equation*}
	\Psi_2^{\alpha(4)}=e_3e_2e_1e_1e_3e_2e_1e_1\psi_2^{\alpha(4)}=192(y_{-,1}^\alpha y_{+,1}^\alpha)^2-192y_{-,1}^\alpha y_{+,1}^\alpha y_{-,2}^\alpha y_{+,2}^\alpha -
\end{equation*}
\begin{equation}
	-576y_{-,1}^\alpha y_{+,1}^\alpha y_{-,3}^\alpha y_{+,3}^\alpha
	+ 192 y_{-,2}^\alpha y_{+,2}^\alpha y_{-,3}^\alpha y_{+,3}^\alpha + 96(y_{-,3}^\alpha y_{+,3}^\alpha)^2.
\end{equation}
It can be checked straightforwardly that $\Psi_2^{\alpha(4)}$ also belongs to
$H^0(\sigma_-)$.

Thus, checking  all possible combinations of the raising operators acting on the vacuum solutions $(y_{+,2}^\alpha)^n$, new functions, annihilated by Cartan operators, can be found,  associated with the off-shell scalar field\,,
\begin{equation}
	h_1 F(e_1,e_2,e_3)(y_{+,2}^\alpha)^{n} = h_2 F(e_1,e_2,e_3)(y_{+,2}^\alpha)^{n} = h_3 F(e_1,e_2,e_3)(y_{+,2}^\alpha)^{n} = 0\,.
\end{equation}

\section{$H^1(\sigma_-)$} \label{sec5}
In the case of $H^1(\sigma_-)$ we  consider the lowest weight sector following
\cite{Gelfond:2013lba}, where this problem has been solved.
Note that in \cite{Gelfond:2013lba}   $\tau_{mn}$ were representing  algebra $o(6)$ rather than  $o(3,3)$ as in this paper. However, for the
 finite-dimensional modules,
the representation theory of orthogonal Lie algebras is equivalent (up to possible (anti)selfduality conditions irrelevant to the analysis of this paper)
to that  of the pseudo-orthogonal ones.   This allows us to use  the results of \cite{Gelfond:2013lba} directly. In \cite{Gelfond:2013lba}, the general case of $sp(2M)$ and $o(r)$ was considered while in our problem we confine ourselves to $M=2$, implying $d=3$,
and $r=6$, implying that the current is trilinear in the off-shell scalars.

 The problem is to solve the homotopy equation (\ref{homotopy}) with the homotopy operator
 (\ref{De}) expressed as a sum of quadratic Casimirs of $o(r)$ and $gl(M)$.
  The representation is described in terms of the  $gl(M)$ Young diagrams (YD) which
  determine the symmetry of the l.h.s.'s of  equations of motion and dynamical fields. These $gl(M)$ YD are associated with  traceless tensors with respect to the color indices $a,b$ in (\ref{color}) that have the same symmetry properties as the $gl(M)$ YD. We  say that the respective traceless symmetric $o(r)$ tensors are described by the traceless symmetric YD.

  In \cite{Gelfond:2013lba} it was shown that $H^1(\sigma_-)$ consists of the traceless
  tensors described by the YD ${\bf Y}_1(n_1,\dots,n_m,\underbrace{2,\dots,2}_{r+1-2m-q},\underbrace{1,\dots,1}_{q})$, where $2m + q \leqslant r$. In our case of $r=6$ these are
  	
  		\vspace{0.3 cm}
  	1) m=0, q=0\dots6, \;\;\;\;\;\;
  	{\ytableausetup{boxsize=1em}
  		\ydiagram{2,2,2,2,2,2,2} \;\;\;\;\;\;  \ydiagram{2,2,2,2,2,2,1}  \;\;\;\;\;\;  \ydiagram{2,2,2,2,2,1,1} \;\;\;\;\;\;  \dots \;\;\;\;\;\;  \ydiagram{2,2,1,1,1,1,1} \;\;\;\;\;\;  \ydiagram{2,1,1,1,1,1,1}\;,
  	}
  	
  	\vspace{0.3 cm}			
  	2) m=1, q=0\dots4, \;\;\;\;\;\;
  	{
  		\ydiagram{6,2,2,2,2,2} \;\;\;\;\;\;  \ydiagram{6,2,2,2,2,1}  \;\;\;\;\;\; \dots \;\;\;\;\;  \ydiagram{6,2,1,1,1,1}\;,
  	}
  	
  	\vspace{0.3 cm}
  	3) m=2, q=0,1,2, \;\;\;\;\;\;\;
  	{
  		\ydiagram{6,4,2,2,2} \;\;\;\;\;\;  \ydiagram{6,4,2,2,1}  \;\;\;\;\;\;  \ydiagram{6,4,2,1,1}\;,
  	}
  	
  	\vspace{0.3 cm}		
  	4) m=3, q=0, \;\;\;\;\;\;\;\;\;\;\;\;\;\; \ydiagram{9,7,5,2} \;.
  	\\
  	
According to the two-column theorem (see, {\it e.g.}, \cite{hamermesh1989group}) traceless tensors corresponding to Young diagrams, in which the sum of the heights of the first two columns exceeds $r$, vanish. This implies that all these diagrams are trivial, which manifests the comment from \cite{Gelfond:2013lba} that for $M<r$ many diagrams trivialize.
In turn, this implies that, in the case of interest, $H^1(\sigma_- )=0$.
Therefore, there are no
non-trivial equations obeyed by the trilinear combinations of the off-shell scalar fields.

Naively, this contradicts the results of \cite{Lavrov}. Before explaining
the origin of this seeming disagreement in Section \ref{sec6} let us make the following
comment. It is easily observed that such YD are trivial also for higher ranks ($r>6$), which stands for arbitrary nonlinear combinations of currents. Indeed, the sum of lengths of the first two columns $l$ of diagrams ${\bf Y}_1(n_1,\dots,n_m,\underbrace{2,\dots,2}_{r+1-2m-q},\underbrace{1,\dots,1}_{q})$ is $l=2r+2-2m-q$. Taking into account that $2m+q \leqslant r$ it is obvious that $l > r$. Thus, all these diagrams and corresponding equations on the multilinear composite fields must be trivial. It also should be mentioned that $H^1(\sigma_-)$ describes the equations of motion for fields of arbitrary spin. According to \cite{Gelfond:2003vh}, the subsystems associated with different eigenvalues of operators $h_i$ (\ref{H}) describe conserved currents of different spins. That $H^1(\sigma_-)$ is empty means that actually there are no dynamical equations not only for multilinear combinations of scalar currents, but also for multilinear combinations of spin-$s$ conserved currents in three dimensions.

\section{Spin three example} \label{sec6}
Thus it is shown that composite fields in question obey
no differential equations (to single out the current of definite spin, for instance spin 3, one has to restrict $H^0(\sigma_-)$ to the polynomials of the corresponding degree in $y_{\pm,i}^\alpha$). On the other hand, the  current (\ref{current}) found in
\cite{Lavrov} does obey the conservation condition.
This seeming contradiction can be resolved however if the current (\ref{current}) is actually deformationally  trivial
in which case the action term (\ref{vert})  can be compensated by a local field redefinition
of the free  Fronsdal action.

The spin-three Fronsdal tensor is \cite{Fronsdal}:
\begin{equation}
\label{freq}
	{\cal R}_{\mu \nu \lambda}[\varphi]   = \Box \varphi_{\mu \nu \lambda} - 3 \partial_{(\mu} \partial^\alpha \varphi_{\nu \lambda) \alpha} + 3 \partial_{(\mu} \partial_{\nu} \varphi_{\;\; \lambda) \alpha}^\alpha.
\end{equation}
By plugging
\begin{equation}\label{dec}
 T^{dec}_{\mu \nu \lambda}:=
\eta_{(\mu \nu} \partial_{\lambda )}\phi \; \phi^2 = \partial_\rho \left( \frac{1}{3} \eta_{(\mu \nu} \delta^\rho_{\lambda)} \phi^3 \right)
\end{equation}
into this expression  we get the  tensor $T_{\mu \nu \lambda}$ (\ref{current}):
\begin{equation}
\label{RT}
	{\cal R}_{\mu \nu \lambda}[T^{dec}] =(-2) T_{\mu \nu \lambda}. \end{equation}
Note that $T^{dec}_{\mu \nu \lambda}$ (\ref{dec}) is a descendant according to the terminology of Section \ref{sec1}.

This implies that there exists such a local transformation of the spin-three field $\varphi_{\mu \nu \lambda}$ that eliminates the vertex (\ref{vert}). Indeed, by substituting
\begin{equation}\label{eq}
\varphi_{\mu \nu \lambda} \to
\varphi_{\mu \nu \lambda} + \delta \varphi_{\mu \nu \lambda}
\end{equation}
 into the Fronsdal action
\begin{equation}
	S[\varphi] := \frac{(-1)^{s+1}}{2} \int d^d x ( \varphi^{\mu_1 \dots \mu_s} {\cal R}_{\mu_1 \dots \mu_s} (\varphi)
	-\frac{s(s-1)}{4} \varphi_\nu^{\;\;\; \nu \mu_3 \dots \mu_s} {\cal R}^\beta_{\;\;\; \beta \mu_3 \dots \mu_s}(\varphi) )\,
\end{equation}
and using that, as one can easily check,  it is a symmetric functional,
\begin{equation}
	\int d^dx \left( \delta \varphi^{\mu_1 \dots \mu_s} {\cal R}_{\mu_1 \dots \mu_s}(\varphi) \right) = \int d^d x \left( \varphi^{\mu_1 \dots \mu_s} {\cal R}_{\mu_1 \dots \mu_s}(\delta \varphi) \right) \,,
\end{equation}
one observes  that the substitution (\ref{eq}) induces the vertex (\ref{vert}).

Clearly, there exists an infinite number of such transformations and corresponding nonlinear conserved currents, because an arbitrary  combination of derivatives and fields $P_{\alpha_1 \alpha_2 \dots}[\partial_\alpha; \phi]$ would produce a current in the same way as for $T^{dec}_{\mu \nu \lambda}$. Being removable by  local field
redefinitions all such vertices are dynamically trivial, however, which conclusion is in
agreement with the results of \cite{Fredenhagen:2019hvb, Joung:2019wbl,Fredenhagen:2019lsz}.

Also let us stress that the resulting currents, including that of (\ref{RT}), are of the form
$L_F(\partial) T^{pr}$ discussed in Introduction, where
$L_F(\partial) $ is the Fronsdal operator on the RHS of
(\ref{freq}).

\section{Conclusion} \label{sec7}
In this paper, interaction vertices of spin-$s$ massless fields and composite fields identified
with nonlinear combinations of  other fields $\phi$, that may obey no field equations thus being off-shell, are considered. It is illustrated how the  $\sigma_-$-cohomology technique can be applied to the analysis of the interaction vertices. To this end, composite fields are interpreted as primaries of the tensor products of the unfolded modules associated with
$\phi$. It is explained that, in general, there may be two types of vertices. The
deformationally trivial ones, that do not demand the deformation of the transformation law,
are associated with the descendants of the unfolded vertex modules. The non-trivial vertices,
associated in particular with usual conserved currents, are in the $\sigma_-$-cohomology.

For simplicity we focus on the case of three dimensions, though some general conclusions are made for any dimension. The $\sigma_-$-cohomology technique, which allows one to control the dynamical sector of the theory in question, was applied to find
out whether composite fields obey any non-trivial equations of motion as well as the nontriviality of their interaction vertices. As  explained in Section \ref{sec3} the problem amounts to the solution of the homotopy equation with the restriction (\ref{H}) that the eigenvalues of the operators $h_i$  are zero. The key observation, that reduces  the problem to the  Lie algebra representation theory, is that $h_i$  form the Cartan subalgebra of the $o(3,3)$ Howe dual \cite{Howe} to the space-time symmetry algebra.
It is explained in detail how to calculate $H^0(\sigma_-)$ in order to find the
independent composite fields in the theory. Howe duality plays instrumental role in this analysis.

Using the results of \cite{Gelfond:2013lba} we have shown that the first cohomology is  empty, $H^1(\sigma_-)=0$,
which means that there are no non-trivial equations of motion obeyed by the composite fields. In particular, this implies  that the conserved current of \cite{Lavrov} must be deformationally trivial, that was checked directly in  any dimension. Moreover, that $H^1(\sigma_-)=0$ implies that no non-trivial equations of motion exist for all (not only trilinear) multilinear combinations of spin-$s$ currents with spin-zero current  identified with the off-shell scalar field. Thus, all possible interaction vertices between
spin-$s$ fields and multilinear currents must be deformationally trivial at least in three dimensions. In agreement with the
earlier papers \cite{Fredenhagen:2019hvb, Joung:2019wbl, Fredenhagen:2019lsz},
we anticipate  this result to be true in any dimension. Note however that
our analysis is  free from the conjectures of \cite{Fredenhagen:2019hvb, Joung:2019wbl, Fredenhagen:2019lsz} on the possibility to use certain non-local operators in the proof.
Another comment is that our results go far beyond the specific case of the
conservation conditions related to HS dynamics, stating that primary components of
the  composite combinations of
the off-shell fields and their derivatives  cannot obey any non-trivial partial differential
equations at least in three dimensions.

  \section*{Acknowledgments}
The authors are  grateful to Olga Gelfond for
useful discussions and comments and to Dmitry Ponomarev for
the correspondence. The comments by Anatoly Shabad are also
appreciated. MV is grateful  to Petr Lavrov and Yuri Zinoviev for the
illuminating conversation and wishes to thank
Ofer Aharony, Theoretical High Energy Physics Group of Weizmann Institute of Science
for hospitality during accomplishment of  this work.

\appendix

\section{Appendix} \label{Appendix}

In this appendix we explain in detail how the system (\ref{soe}) can be solved.
Let us examine the first two equations in this system:
\begin{equation}
	\label{low}
	\begin{cases}
		\left( y_{-,1}^\alpha \frac{\partial}{\partial y_{-,3}^\alpha}-y_{+,3}^\alpha\frac{\partial}{\partial y_{+,1}^\alpha} \right) C = 0, \\
		\left( y_{-,1}^\alpha \frac{\partial}{\partial y_{+,3}^\alpha}-y_{-,3}^\alpha \frac{\partial}{\partial y_{+,1}^\alpha} \right) C = 0.\\
	\end{cases}
\end{equation}
We will look for a solution in the form $C=\sum_{n=0} (y_{-,1}^\alpha)^n C_n$. In the zeroth order in $y_{-,1}^\alpha$ we get $\frac{\partial C_0}{\partial y_{+,1}^\alpha} = 0$, that is
\begin{equation}
	C_0=C_0(y_{\pm,3})\,.
\end{equation}
In the first order in $y_{-,1}^\alpha $ we obtain
\begin{equation}
	\begin{cases}
		\frac{\partial}{\partial y_{-,3}^\alpha }C_0 = y_{+,3}^\alpha  \frac{\partial}{\partial y_{+,1}^\alpha }C_1, \\
		\frac{\partial}{\partial y_{+,3}^\alpha }C_0 = y_{-,3}^\alpha  \frac{\partial}{\partial y_{+,1}^\alpha }C_1.
	\end{cases}
\end{equation}
Suppose that $C_0$ can be expressed as $C_0=y_{-,3}^\alpha y_{+,3}^\alpha C_0'(y_{\pm,3})$.
Then
\begin{equation}
	\begin{cases}
		C_0' + y_{-,3}^\alpha \frac{\partial}{\partial y_{-,3}^\alpha }C_0' = \frac{\partial}{\partial y_{+,1}^\alpha}C_1, \\
		C_0' + y_{+,3}^\alpha  \frac{\partial}{\partial y_{+,3}^\alpha }C_0' = \frac{\partial}{\partial y_{+,1}^\alpha }C_1\,.
	\end{cases}
\end{equation}
Since the left-hand side is $y_{+,1}^\alpha$ independent, the equations can be easily integrated
\begin{equation} \label{A1}
	\begin{cases}
		C_1 = y_{+,1}^\alpha (C_0'+y_{-,3}^\alpha \frac{\partial}{\partial y_{-,3}^\alpha }C_0') + c_1(y_{\pm,3}), \\
		C_1 = y_{+,1}^\alpha (C_0'+y_{+,3}^\alpha \frac{\partial}{\partial y_{+,3}^\alpha }C_0') + c_2(y_{\pm,3})\,.
	\end{cases}
\end{equation}
By subtracting the second equation from the first, we get
\begin{equation}
	y_{-,3}^\alpha  \frac{\partial}{\partial y_{-,3}^\alpha }C_0' = y_{+,3}^\alpha  \frac{\partial}{\partial y_{+,3}^\alpha } C_0', \quad c_1 = c_2 = C_0^1(y_{\pm,3})\,.
\end{equation}
This implies
\begin{equation}\label{A2}
	C_0 = \sum_{n}\alpha_n (y_{-,3}^\alpha y_{+,3}^\alpha )^n.
\end{equation}
Plugging \eqref{A2} into \eqref{A1} we obtain
\begin{equation} C_1=y_{+,1}^\alpha C^1_1+C_0^1, \;\;\;\;\; C_1^1=\sum_{n}n\alpha_{n}(y_{-,3}^\alpha y_{+,3}^\alpha )^{n-1},\;\;\;\;\; C_0^1=C_0^1(y_{\pm,3}).
\end{equation}
The equations for other $C_k^n$ are analogous. Finally, the solution can be written in the form of $y$ polynomials:
\begin{equation}
	C_b^m=\sum_{k=0}^{n-m-b} (k+b)_b \alpha_{k,m-b} (y_{-,3}^\alpha y_{+,3}^\alpha )^{k},\;\;\;\;\; (k+b)_b := \prod_{\gamma=0}^{b-1}(k+b-\gamma)=\frac{(k+b)!}{k!}, \;\;\;\;\; C_n=\sum_{b=0}^{n}\frac{(y_{+,1}^\alpha )^b}{b!}C_b^n,
\end{equation}
\begin{equation}\label{A3}
	C=\sum_{b=0}^m\sum_{k=0}^{n-m}\sum_{m=0}^{n}(y_{-,1}^\alpha)^m\frac{(y_{+,1}^\alpha )^b}{b!} (k+b)_b \alpha_{k+b,m-b}(y_{-,3}^\alpha y_{+,3}^\alpha )^{k}.
\end{equation}
Since the equations (\ref{low}) are first-order partial differential equations they can be
solved by the method of characteristics implying that any function of some
solutions is a solution. As is not hard to check straightforwardly,
there are the following two simple solutions
\begin{equation}
	\label{g12}
	g_1^\alpha=y_{-,1}^\alpha, \;\;\; g_2^{\alpha_1 \alpha_2}=y_{-,3}^{\alpha_1} y_{+,3}^{\alpha_2} +y_{-,1}^{\alpha_1} y_{+,1}^{\alpha_2}\,.
\end{equation}
At $n=1$ a solution \eqref{A3} can be written in the form
\begin{equation}
	C=(\alpha_{00}+(\alpha_{10})_{\alpha \alpha}y_{-,3}^\alpha y_{+,3}^\alpha )+y_{-,1}^\alpha((\alpha_{01})_\alpha+y_{+,1}^\alpha (\alpha_{10})_{\alpha \alpha})=\alpha_{00}+(\alpha_{10})_{\alpha \alpha}
	g_2^{\alpha\alpha}+(\alpha_{01})_{\alpha}g_1^\alpha\,.
\end{equation}
It is not hard to make sure by induction that the obtained solution can be expressed
as a polynomial of $g_1$ and $g_2$ at any order. Assuming that this is true at the $n$-th step one can easily check that it is also  true at the ($n+1$)-th step.
Thus it is shown that  the solution (\ref{A3}) is a polynomial of $g_1^\alpha$ and $g_2^{\alpha_1 \alpha_2}$ (\ref{g12}).

 Solutions to the full system of equations (\ref{vac}) can be searched in the form $g_{1,2} + P(y_{\pm 2})$. After some straightforward calculations the elementary solutions to (\ref{vac}) can be written down as:
\begin{equation}
	G_1^\alpha = y_{+,2}^\alpha, \quad G_2^{\alpha \alpha} = y_{+,1}^\alpha y_{-,1}^\alpha + y_{+,2}^\alpha y_{-,2}^\alpha + y_{-,3}^\alpha y_{+,3}^\alpha\,.
\end{equation}

\end{document}